# Experimental demonstration of a crossed cubes nuller for coronagraphy and interferometry

François Hénault, Brahim Arezki, Guillaume Bourdarot
Institut de Planétologie et d'Astrophysique de Grenoble
Université Grenoble-Alpes, Centre National de la Recherche Scientifique
B.P. 53, 38041 Grenoble – France

Alain Spang
Laboratoire Lagrange, Université Côte d'Azur
Observatoire de la Côte d'Azur, CNRS,
Parc Valrose, Bât. H. FIZEAU, 06108 Nice – France

**ABSTRACT**

In this communication we present the first experimental results obtained on the Crossed-cubes nuller (CCN), that is a new type of Achromatic phase shifter (APS) based on a pair of crossed beamsplitter cubes. We review the general principle of the CCN, now restricted to two interferometric outputs for achieving better performance, and describe the experimental apparatus developed in our laboratory. It is cheap, compact, and easy to align. The results demonstrate a high extinction rate in monochromatic light and confirm that the device is insensitive to its polarization state. Finally, the first lessons from the experiment are summarized and discussed in view of future space missions searching for extra-solar planets located in the habitable zone, either based on a coronagraphic telescope or a sparse-aperture nulling interferometer.

## 1 INTRODUCTION

Coronagraphy [1], nulling interferometry [2-3], and transit spectroscopy [4] are three of the most powerful techniques envisaged today for localizing and characterizing extra-solar planets in the habitable zone of their mother stars. The present communication only deals with the two first types of instrumentation. In a recent paper [5] we described a very simple and "cheap" Achromatic phase shifter (APS), that is the core sub-system of a nulling interferometer and an excellent candidate for generating central extinction in coronagraph telescopes. This is the Crossed-cubes nuller (CCN) made of a couple of crossed beamsplitter cubes and of multi-axial focusing optics (see Figure 1). In particular it was shown that the device is suitable for both nulling interferometry and coronagraphy and offers important advantages such as high throughput, very narrow Inner working angles (IWA), and capacities of fringes rotation and modulation. Moreover it is compact, of reasonable manufacturing tolerances, and potentially inexpensive. Detailed performance evaluation and discussion were also provided in Ref. [5].

The purpose of this communication is twofold: Firstly the theoretical principle of the CCN is reviewed and improved taking into consideration some fundamental properties of symmetric semi-reflective layers, which were not discussed in the previous study (section 2). Secondly, we present the first experimental results obtained in our laboratory, including the influence of polarization and spectral bandwidth (section 3). Finally a brief conclusion about the potential and future developments of the CCN is drawn in section 4.

## 2 PRINCIPLE

In section 2.1 we recall the basic optical setup of the CCN as it was described in Ref. [5]. We then recall some elementary properties of symmetric beamsplitters and discuss their dependence vs. wavelength (section 2.2). We demonstrate how this dependence is detrimental to the nulling performance of a classical Mach-Zehnder interferometer (section 2.3). In section 2.4 we present a slightly modified version of the original CCN, and show its advantages with respect to the previous configurations.

### 2.1 Basic CCN setup

The original optical setup of the CCN is illustrated in Figure 1. It is composed of two beam splitting cubes having perpendicular semi-reflective layers and of an achromatic doublet focusing the four output beams onto a common image point O'. The reference frame attached to the CCN is defined as follows:

- o  Z is the main optical axis, parallel to the optical axis of the focusing lens,
- o  X is perpendicular to the Z-axis and to the semi-reflective layer of Cube 1, hence parallel to YZ plane,
- o  Y is perpendicular to the Z-axis and to the semi-reflective layer of Cube 2, hence parallel to XZ plane.

We consider a flat monochromatic wavefront propagating in the +Z direction, and impinging Cube 1 under an incidence angle equal to 45 degrees. The wave is linearly polarized either along the X-axis (thick red lines on the Figure) or along the Y-axis (thin black lines). It is first refracted by the input face of Cube 1, then split in two by the YZ semi-reflective layer. After being refracted again at the output faces of Cube 1, both beams are parallel and directed along the Z-axis, but due to phase shifts occurring at the layer their relative polarization states have changed (note that polarization flips at reflection are equivalent to achromatic π-phase shifts). The same process is then repeated on Cube 2, where phase shifts generated by the XZ semi-reflective layer are added to those of Cube 1. The four parallel output beams exiting Cube 2 are finally combined multi-axially by the achromatic lens, creating an interference pattern in the O'X'Y' image plane. Because these phase shifts are by nature achromatic and polarization-independent, it follows that the CCN can meet all the critical requirements of an APS usable for extra-solar planets detection and spectroscopic characterization, provided that the phase shifts and amplitude balance are carefully selected for generating deep central extinction at the image plane.

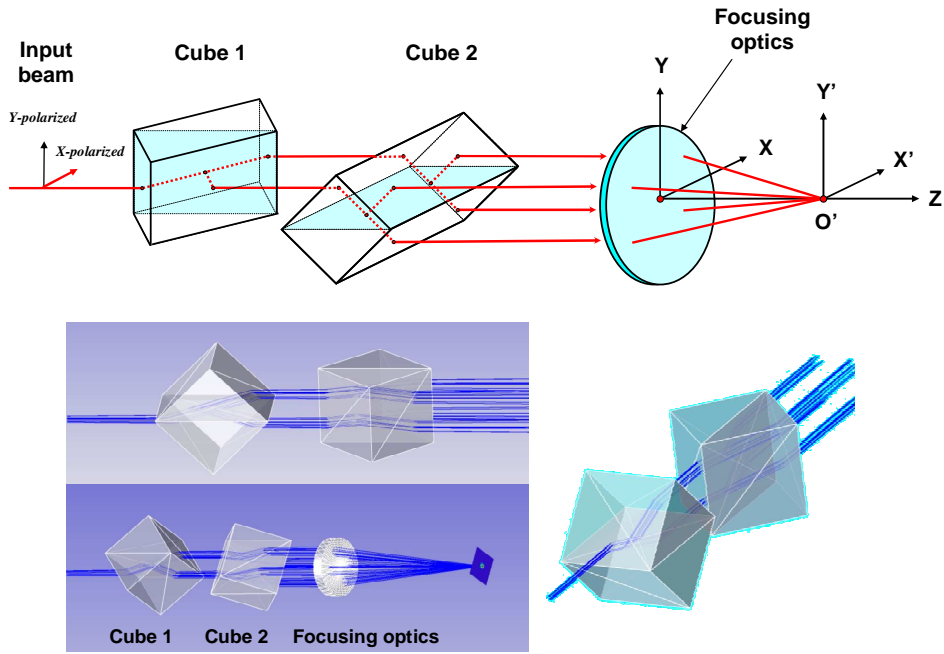

Figure 1: General views of the crossed-cubes nuller (from Ref. [5]).

Many properties and special advantages of the CCN are extensively discussed in Ref. [5]. They are briefly summarized below:

- The CCN can be seen as a three-dimensional version of the classical Mach-Zehnder (MZ) interferometer where the four output ports would be spatially separated. Equivalently, the CCN could be turned into a MZ interferometer with two output ports by setting parallel the semi-reflective layer planes of Cubes 1 and 2.

- This APS can either be integrated into a nulling interferometer array or a monolithic coronagraphic telescope. In both cases, extreme on-axis rejection ratios are achievable.

- When used as coronagraph, it allows easy adjustment of the exit baselines, thus reaching very narrow Inner working angles (IWA).

- Excepting flux equalization, most of the cubes optical manufacturing tolerances are quite moderate. The variant described in this communication allows relaxing the flux balance requirements. Therefore the cubes and the focusing lens could be of commercial grade.

The most important requirements of the splitting cubes is that they are symmetric and lossless, so that their amplitude reflection and transmission coefficients $A_R$ and $A_T$ have identical square modules $|A_R|^2 = |A_T|^2 = 0.5$. It also has important consequences on the values of the phase-shifts $\phi_R$ and $\phi_T$, respectively occurring after reflection or transmission at the semi-reflective layers. This point is developed in the next sub-section.

## 2.2 Elementary properties of symmetric beamsplitters

Evaluating the phase-shifts $\phi_R$ and $\phi_T$ of the beams reflected and transmitted by a symmetric beam-splitting cube[1] is not extremely complicated, but deserves a few words of explanation. This topic was the scope of a recent communication [6], though written in a slightly different context (comparing the mathematical formalisms of physical optics and quantum optics, and their physical interpretation). Applying them to the CCN, the major results of this paper are illustrated in Figure 2 and summarized below. First the complex amplitudes $A_R$ and $A_T$ of the reflected and transmitted beams are computed as the two output sums of a multi-interference effect occurring inside the semi-reflective layer of the cube (this is the same treatment as for a Fabry-Perot interferometer). For the case of a lossless semi-reflective layer that is assumed here, it leads to[2]:

$$A_T = T \frac{\exp(i\varphi)}{1 - R\exp(2i\varphi)} \quad (1a) \qquad \text{and:} \qquad A_R = -\sqrt{R} \frac{1 - \exp(2i\varphi)}{1 - R\exp(2i\varphi)}, \quad (1b)$$

where $\varphi$ is the internal phase-shift of the layer: $\quad \varphi = 2\pi n e \cos\theta / \lambda$, (2)

and the following notations are used (see Figure 2, panel A):

| | |
|---|---|
| $T$ | Intensity transmission coefficient at glass/layer interface |
| $R$ | Intensity reflection coefficient at glass/layer interface |
| $n$ | refractive index of the layer, |
| $e$ | thickness of the layer, |
| $\theta$ | internal incidence angle at layer interface, |
| $\lambda$ | wavelength of the electro-magnetic field. |

Two important consequences may be derived from the here above equations 1:

1) Combining Eqs. 1 readily leads to the necessary conditions for perfect flux balance $|A_R|^2 = |A_T|^2 = 0.5$, that is:

$$\varphi = \pm \arcsin\left(T/2\sqrt{R}\right) \; [2\pi]. \quad \text{or} \quad \varphi = \pi \mp \arcsin\left(T/2\sqrt{R}\right) \; [2\pi], \quad (3)$$

or equivalently in terms of the layer thickness (see Eq. 2) and assuming that $T = R = 0.5$:

---

[1] Here and in the remainder of the study are neglected the effects of a thin glue layer deposited on the semi-reflective layer for cube assembly.
[2] Using a common reference point for both the transmitted and reflected beams as explained in Ref. [6], § 2.2.

$$e = (k \pm 0.0575)\lambda/n\cos\theta \quad \text{or} \quad e = (k + 0.5 \mp 0.0575)\lambda/n\cos\theta, \tag{4}$$

where $k$ is an integer number. Thus over any $[0,2\pi]$ interval it is possible to define four different values of $\varphi$ (or equivalently four layer thickness $e$) ensuring equal reflected and transmitted intensities, as illustrated in Figure 2, panel B.

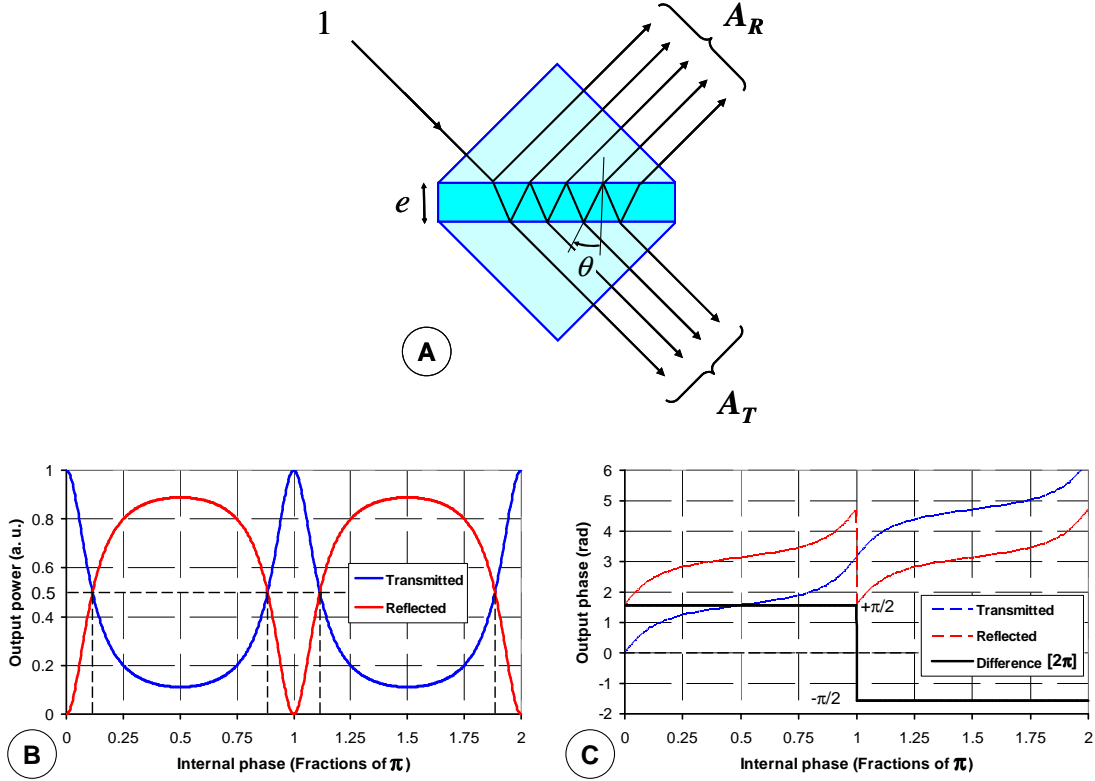

Figure 2: Basic model of symmetric beam-splitting cubes. A) Multi-interference representation. B) Reflected and transmitted intensities $|A_R|^2$ and $|A_T|^2$. C) Phase-shifts $\phi_R$ and $\phi_T$ of the reflected and transmitted beams. It is assumed that $T = R = 0.5$.

2) The phase-shift between the reflected and transmitted beams $\phi_{R1} - \phi_{T1}$ is also obtained easily by determining the argument of the complex quantity $A_R A_T^*$. It is found that:

$$\phi_R - \phi_T = Arg[A_R A_T^*] = Arg[i\sin\varphi] = \pm\frac{\pi}{2}. \tag{5}$$

This result demonstrates that the phase-shift is always equal to one-fourth of the impinging wave regardless of other optical phenomena such as wavelength-dependence of the material or polarization of the electromagnetic field. This fundamental independence is illustrated in the panel C of Figure 2.

From this very short study, it can be concluded that there exist at least two different families of cube beamsplitters, the first one generating an optical advance of $\pi/2$ to the transmitted beam with respect to the reflected beam, while the second one delays the transmitted beam by $\pi/2$. Practically speaking, it is likely that cube beamsplitters provided by a given manufacturer will all be from the same family, as imposed by the original optical specifications (thickness and material of the semi-reflective layer, most cubes being probably made of a single one). The uncertainty on the sign of $\phi_R - \phi_T$ can only be removed if the optical prescription of the coating is available.

Before returning to the CCN, it is worth studying the caser of a Mach-Zehnder interferometer made of two symmetric cube beamsplitters, that is the purpose of the following sub-section.

**2.3 The Mach-Zehnder interferometer (MZ)**

The Mach-Zehnder (MZ) is a well-known type of interferometer originally conceived for comparing two parallel and collimated optical beams by means of their phase difference. It has been successfully employed either as a nulling coronagraph [7]; or as combining unit for the PERSEE nulling interferometer [8] – where was measured a $10^{-5}$ nulling ratio over a 37 % spectral bandwidth. The case when it is equipped with two symmetric beamsplitter cubes is less common in literature, although one may cite Refs. [9-10]. It is illustrated in the upper panel of Figure 3, showing its general configuration and the output intensities and phase shifts for the (arbitrarily chosen) case when $\phi_R - \phi_T = -\pi/2$. In the lower panel of the figure are indicated the resulting phase shifts in both output ports of the interferometer, labeled 1 and 2 for the constructive output port and destructive, "null" port respectively.

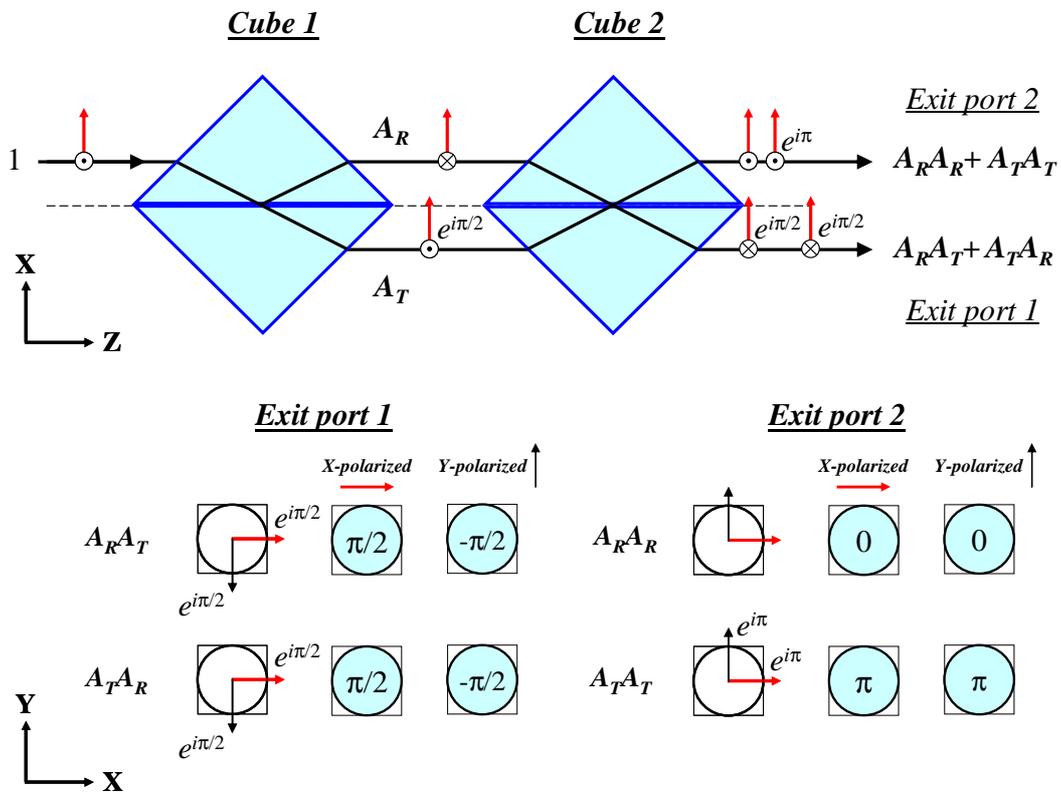

Figure 3: General scheme of a MZ interferometer equipped with beam splitting cubes, and of its output intensities and phase shifts. The latter are indicated for each sub-pupil and linear polarization direction.

Employing the same notations as in sub-section 2.2, the expressions of the amplitudes and intensities at the output ports 1 and 2 of the MZ interferometer are written classically:

|  | Complex amplitude | Intensity |  |
|---|---|---|---|
| Output port 1 (constructive) | $A_1 = 2 A_T A_R$ | $I_1 = |A_1|^2 = 4 |A_T|^2 |A_R|^2$ | (6a) |
| Output port 2 (null) | $A_2 = A_T A_T + A_R A_R$ | $I_2 = |A_2|^2 = |A_T A_T + A_R A_R|^2$ | (6b) |

From Eqs. 1 we first compute the transmitted and reflected intensities of a single cube beamsplitter:

$$|A_T|^2 = \frac{T^2}{1+R^2-2R\cos 2\varphi} \quad (7a) \quad \text{and:} \quad |A_R|^2 = \frac{4R\sin^2\varphi}{1+R^2-2R\cos 2\varphi}. \quad (7b)$$

Combining Eqs. 1 and 7 and applying the principle of energy conservation at the exit of the MZ interferometer $|I_1|^2 + |I_2|^2 = 1$ then leads to the following mathematical expressions of the output intensities $I_1$ and $I_2$:

$$I_1 = \frac{16\,RT^2\sin^2\varphi}{\left(T^2+4R\sin^2\varphi\right)^2} \quad (8a) \quad \text{and:} \quad I_2 = \frac{T^4-8RT^2\sin^2\varphi+16R^2\sin^4\varphi}{\left(T^2+4R\sin^2\varphi\right)^2}. \quad (8b)$$

From Eqs. 3 the necessary condition for flux balance is $\sin^2\varphi = T^2/4R$, yielding $I_1 = 1$ and $I_2 = 0$ as expected for the MZ interferometer with port 1 being constructive and port 2 being destructive. One may realize however that Eqs. 3 are only valid at a given reference wavelength $\lambda_0$, part of the optical prescription of cube beamsplitter. In reality T and R will be functions of the wavelength $\lambda$ and Eqs. 8 shall be written more precisely:

$$I_1(\lambda) = \frac{4T^2(\lambda_0)}{R(\lambda_0)}\,\frac{R(\lambda)T^2(\lambda)}{\left(T^2(\lambda)+R(\lambda)T^2(\lambda_0)/R(\lambda_0)\right)^2} \quad (9a)$$

$$\text{and:} \quad I_2(\lambda) = \frac{T^4(\lambda)-2R(\lambda)T^2(\lambda)T^2(\lambda_0)/R(\lambda_0)+R^2(\lambda)T^4(\lambda_0)/R^2(\lambda_0)}{\left(T^2(\lambda)+R(\lambda)T^2(\lambda_0)/R(\lambda_0)\right)^2}. \quad (9b)$$

Hence the nulled intensity $I_2$ will depend on $\lambda$ and only be equal to zero when $\lambda = \lambda_0$. These last relations illustrate the intrinsic drawback of the MZ interferometer employed as a nuller on its destructive port: there remains a chromatic flux unbalance degrading the extinction ratio on large spectral bandwidths. It will be seen in the next sub-section that a modified version of the CCN setup allows overcoming this fundamental limitation.

### 2.4 Modified CCN setup

The modified optical setup of the CCN is illustrated in Figure 4. It is indeed very similar to its original description (Ref. [5], Fig. 2), the sole difference residing in the fact that actual achromatic phase shifts of $\pi/2$ generated by the beam splitting cubes (as defined in § 2.2) have been introduced into the model (previously assuming $\pi$-phase-shifts, which is untrue for symmetric beamsplitters). The resulting output sub-pupil maps are shown on the central panel of the figure, indicating their different amplitudes and phase-shifts for two input linear polarization states (thick red lines along the X-axis and thin black lines along the Y-axis). It can be seen that they do not meet the elementary condition for generating a central null with N sub-apertures (N=4 in our case), that is:

$$\sum_{n=1}^{N}\exp(i\varphi_n)=0\,, \quad (10)$$

where $\varphi_n$ are the individual phase-shifts in each output sub-pupil. This is due to the fact that the output beams $A_T A_T$ and $A_R A_R$ (respectively transmitted and reflected at both cubes) show identical phase-shifts whatever the polarization state. We see however that the two other beams $A_R A_T$ (reflected by Cube 1 then transmitted by Cube 2) and $A_T A_R$ (transmitted by Cube 1 then reflected by Cube 2) still fulfill the necessary condition 10. It should then be sufficient to remove the two beams $A_T A_T$ and $A_R A_R$ from the final interference plane (for example by means of a diaphragm as sketched on the bottom panel of Figure 4, or of small folding mirrors redirecting the beams towards the metrology sensor) to achieve favourable conditions for deep nulling. Moreover in that case the wavelength dependence of $A_T$ and $A_R$ automatically cancels since only the terms $A_R A_T$ and $A_T A_R$ are made to interfere destructively[1]. In particular, it allows removing the fundamental limitation of the Mach-Zehnder interferometer mentioned in the previous sub-section. Hence this improved CCN setup is not sensitive to polarization or to any chromatic effect of semi-reflective layers, including flux unbalance.

---

[1] Here one may see a pure topological phase-shift of $\pi$, only depending on geometrical transformations and not on the physical properties of materials.

Despite this slight modification of the original design, most of the unique properties of the CCN discussed in Ref. [5] remain valid, as well as its manufacturing tolerances. Significant changes with respect to the original paper are highlighted below:

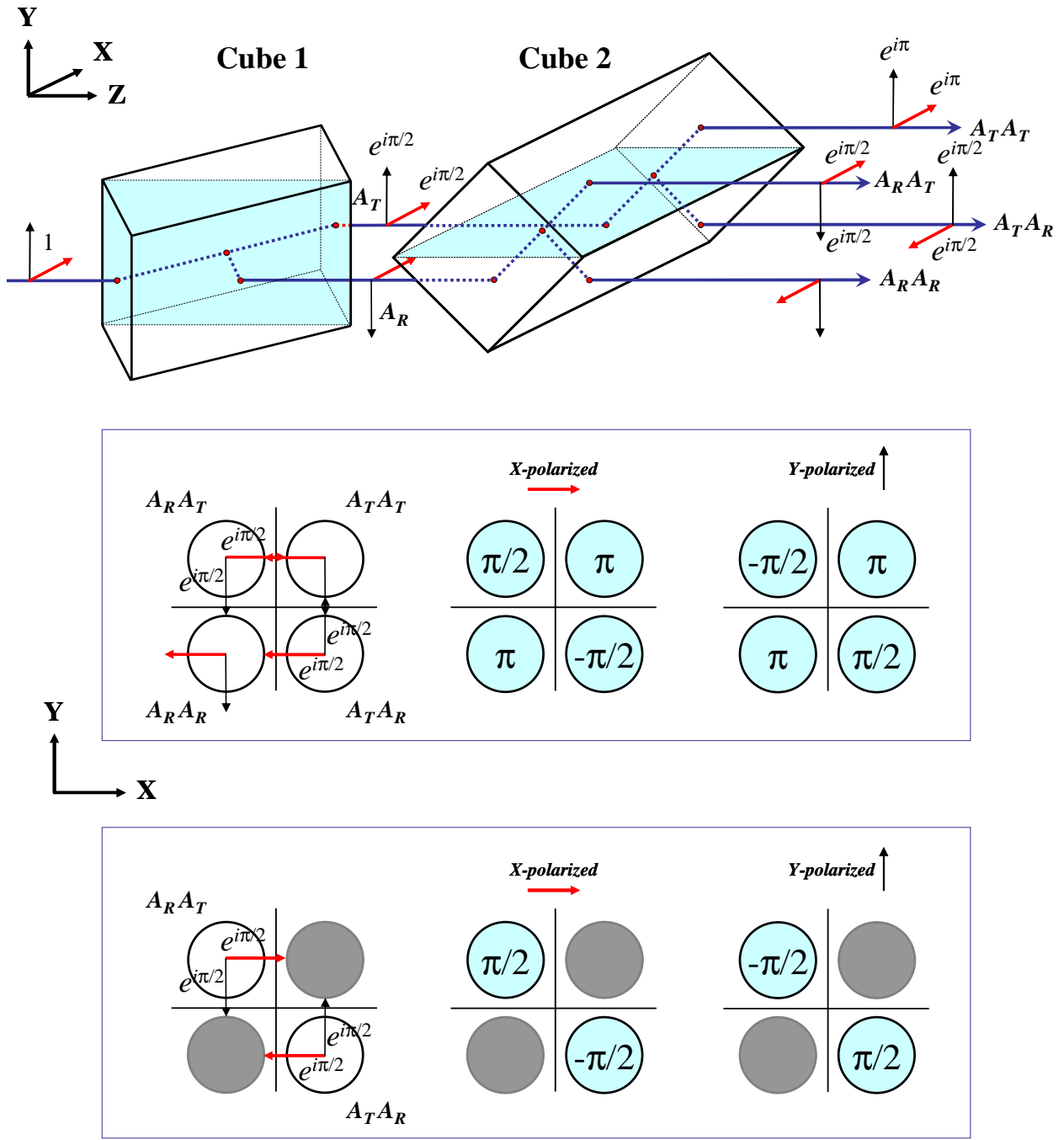

Figure 4: Detailed view of the modified CCN configuration (upper panel), illustrating the different phase-shifts and polarization state of the four split beams (bottom panels).

1) **Nulling maps**: The nulling or transmission maps are of the most fundamental characteristic of a star-occulting instrument. They are defined as the normalized flux amount emitted from or reflected by an extra-solar planet and collected by the system, as a function of the star-planet angular separation [11-12]. Because they originate from only two sub-pupils, the nulling maps produced by the here modified CCN obviously differ from those presented in Ref. [5] (§ 4.1, Fig. 7), though are computed in the same manner. Here the nulling maps shall appear as classical Young's fringes oriented at 45 degrees from the X'-axis, and are independent from the polarization state as will be seen in section 3.

   It must be emphasized that our modified CCN setup can still be used for operating with four sub-pupils as in Ref. [5]. For that purpose one may for example add an optical path length compensator located between Cube 2 an the focusing lens, designed for generating an achromatic $\pi$-phase-shift on the $A_T A_T$ beam (+X+Y sub-pupil in the central panel of Figure 4). Practically this can be achieved by means of a four-arm dioptric compensator, each arm incorporating a couple of wedge plates as described in Ref. [5] (§ 4.2.2, Fig. 8).

2) **Accommodation into coronagraph or nulling interferometer**: As already pointed out the principle of the CCN is suitable to both nulling interferometry and coronagraphy. The implementation schemes of Fig. 3 and Fig. 5 in Ref. [5], respectively into a visible nulling coronagraph or as the Achromatic phase shifter (APS) of an infrared nulling interferometer remain valid here. The modified version offers an interesting possibility to pick the unused $A_T A_T$ and $A_R A_R$ beams for metrology operation (tip-tilt and wavefront sensing, fringe tracking).

3) **Inner working angle** (IWA): When used as coronagraph, the CCN potentially offers very small IWAs well below the critical $\lambda/D$ limit. This advantage is present in the modified version, the theoretical relationship (2) in Ref. [5] remaining fully applicable.

4) **Baseline modulation**: Also in the coronagraph case, the ability of dynamically modulating the fringe pattern is preserved. It not only concerns the inter-fringe distance, but also a limited capacity of rotating the fringe pattern around the optical axis as long as the exit sub-pupils do not overlap.

5) **Manufacturing tolerances**: Finally, most of the previous manufacturing requirements of the beam splitting cubes defined in Ref. [5] (§ 4.2.3, Table 1) remain applicable. One of them however shall be dramatically relaxed: this is the flux mismatch or equalization specification originally set to 0.1 % in absolute value over full operational spectral band. The fact that the complex amplitudes interfering destructively are now identical (i.e. $A_R A_T$ and $A_T A_R$) allows transforming this absolute requirement into a similarity requirement between Cube 1 and Cube 2 (the absolute specification itself could be arbitrarily set to ±5 %). Such a similarity requirement looks quite reasonable if both cubes are issued from the same manufacturing batch. This is undoubtedly the strongest advantage of the modified CCN setup presented here.

## 3   EXPERIMENT AND RESULTS

In this section are briefly described a laboratory experiment developed to test the principle of the modified CCN (§ 3.1) and the first achieved experimental results (§ 3.2 and 3.3).

### 3.1   Experimental setup

In order to confirm the basic principle of the CCN and to have a first estimation of its achievable performance, a first laboratory experiment was built in the Institut de Planétologie et d'Astrophysique de Grenoble (IPAG). In Figure 5 are shown two pictures of our experimental setup, essentially illustrating the two beam splitting cubes and their mounts. Cube 1 and Cube 2 are BK7 made, 1 inch side each. Four of them were ordered from Thorlabs (50:50 non-polarizing beamsplitter cube, ref. BS013) in order to select the most similar couple and insert it into the experiment. The cubes ate mounted on two 2-inch tip-tilt and rotation stages from Newport. The cubes are set perpendicular by means of a in-house made square bracket. The employed light source is a T-cube laser diode at 635 nm (ref. TLS001 from Thorlabs). Its output beam is collimated by an off-axis parabola and limited by a small circular stop (typical diameter 1-2 mm) before impinging Cube 1. The cubes are coarsely positioned so that the exit baselines of the output beams are all around 2 mm.

The focusing lens is from Melles-Griot, 100-mm focal length. The fringes formed in the X'Y' plane are imaged on a digital camera from Thorlabs (ref. DCC1645C) through a x10 microscope objective. In principle this CMOS camera has a dynamic range of 68.2 dB, which is far sufficient for our purpose.

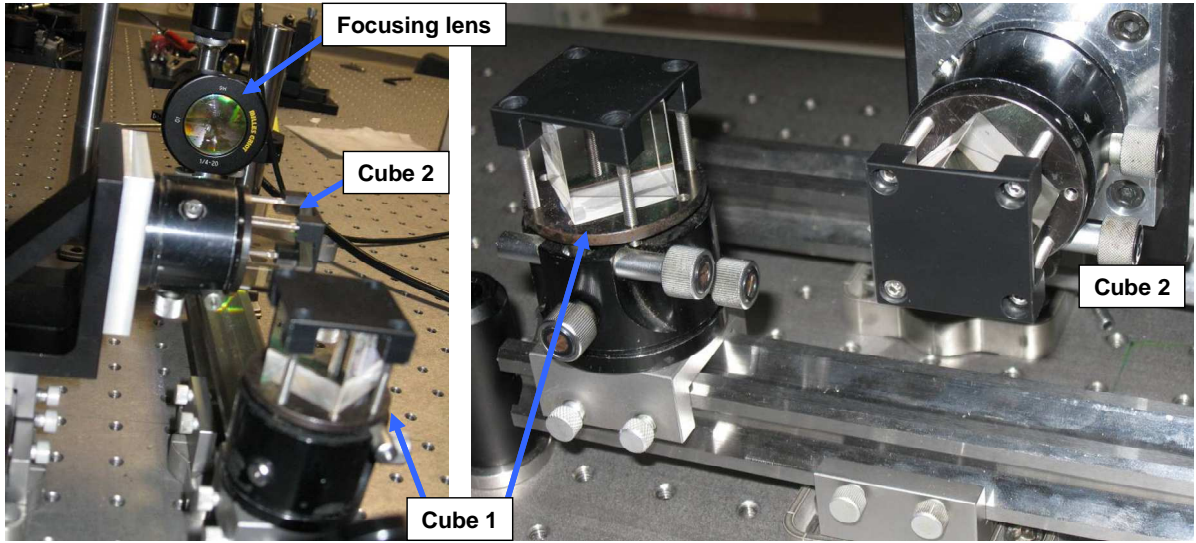

Figure 5: Two pictures of the CCN experimental setup.

### 3.2   Results (first set)

The best results obtained in a first attempt are reproduced in the Figure 6. From the left to the right columns, it shows:

- Case 1: Interference fringes and their power spectral density, obtained when setting a diaphragm with two diagonal holes between Cube 2 and the focusing lens in order to select the $A_R A_T$ and $A_T A_R$ beams and to block the two other beams.

- Case 2: Interference fringes and power spectral density generated from the $A_R A_T$ and $A_T A_R$ beams, obtained when turning over the exit diaphragm.

- Case 3: Fringes and power spectral density generated from the four exit sup-pupils (no diaphragm).

The achieved null ratios are classically evaluated as $I_{Min}/I_{Max}$, i.e. the ratio of minimal and maximal intensities measured over a limited area at the centre of the fringe pattern (shown by yellow boxes in the figure). In this first attempt, they were found to be equal to $6.6 \ 10^{-3}$ (1/150.6) and $9.8 \ 10^{-2}$ (1/10.2) for cases 1 and 2 respectively. Since all the images were digitized over 256 grey levels, it follows that the null ratio measured for case 1 is very close to the noise limit (here arbitrarily set to zero). These results are also particularly interesting because they allow estimating the potential gain in performance resulting from the destructive interference of two perfectly symmetric beams. Here the gain can be estimated as a factor 15. A comparable gain is also to be expected when comparing the CCN with the classical Mach-Zehnder interferometer used in its destructive port, such as in Refs. [7-8].

It is also of interest studying the power spectral densities of the interferograms displayed on the bottom panels of Figure 6. In addition to the expected symmetrical lateral peaks at the angular frequencies $f'_X = \pm B'_X/\lambda$ and $f'_Y = \pm B'_Y/\lambda$ (shown by yellow circles), where $B'_X$ and $B'_Y$ are the exit baselines of the CCN, one can recognize the presence of a number of parasitic frequencies, denoting the existence of spurious interferograms. Their origin is not clearly understood currently. We note however that the parasitic peaks do not generally correspond to pure harmonics of the fundamental frequencies $f'_X$ and $f'_Y$. The origin of those parasitic interferograms could be stray reflections inside the cubes or the focusing lens, whose anti-reflective coatings are of standard class.

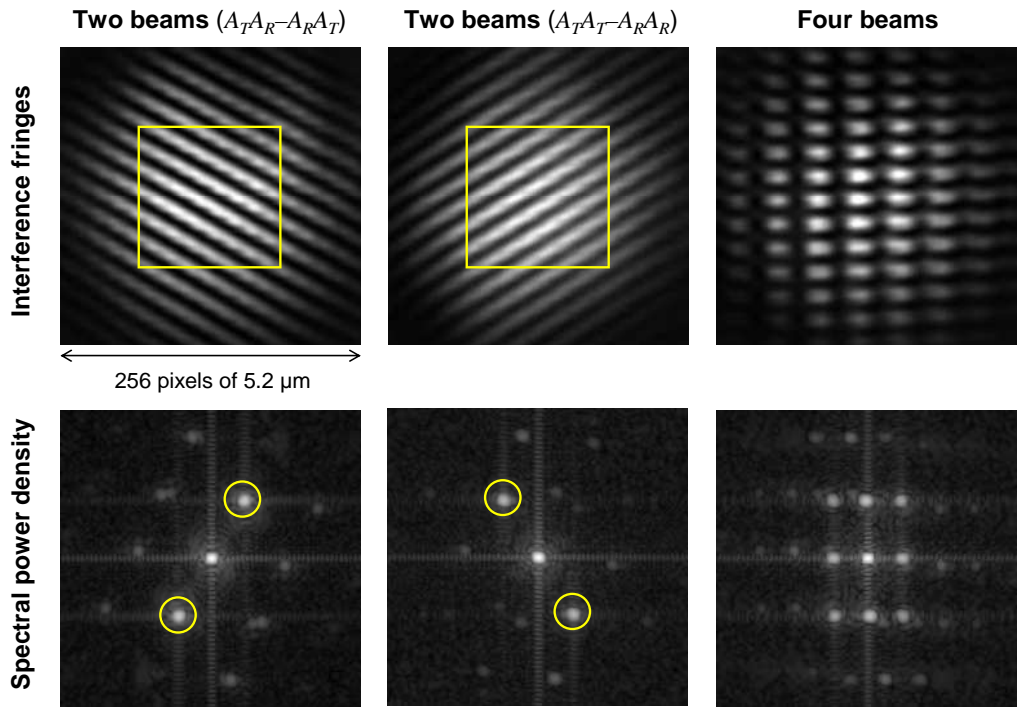

Figure 6: Measured interference fringes generated by the CCN and their power spectral density for the cases of two diagonal interfering beams $A_R A_T$ and $A_T A_R$ (left column), two diagonal interfering beams $A_T A_T$ and $A_R A_R$ (central column), and all four beams interfering (right column).

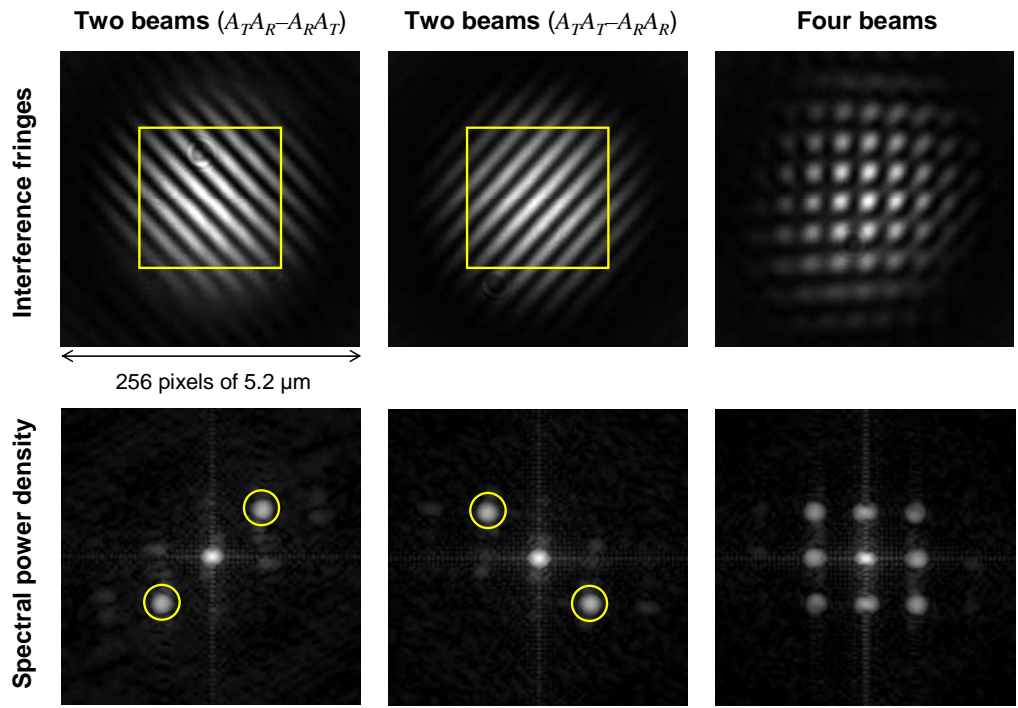

Figure 7: Same illustration as in Figure 6 for our second set of measurements.

### 3.3 Results (second set)

A second set of measurements was performed after careful realignment of the laser diode fiber at the focus of the off-axis parabola with the help of a shear plate tester. The resulting gain in image quality seriously improved the null performance (it must be recalled that in most of nulling experiments[1] the wavefront errors are filtered by single-mode waveguides located after the final beam combination, which is not the case in our experiment). Some of the acquired interferograms are reproduced in Figure 7. The best achieved null ratios are now equal to $1.8 \, 10^{-3}$ (1/554) and $2.2 \, 10^{-3}$ (1/450) for cases 1 and 2 respectively. These last results are very promising since they are below the digitization and readout noises levels of our camera. In the mean time, the number and amplitudes of parasitic interferograms decreased, as can be seen on the power spectral densities (bottom row of the figure). One can finally note the presence of a few specks of dust on the detector array, who should not affect significantly the measured nulling rates since they are located at the edges of the fringe patterns.

## 4   CONCLUSION

In this paper we have reviewed and improved the Crossed-cubes nuller (CCN) concept, recently proposed for the detection of extra-solar planets either in the frame of coronagraphy or nulling interferometry. In particular, a more realistic description of the achromatic phase-shifts occurring at both transmission and reflection by the symmetric beam splitting cubes was given. It allows selecting the most favorable configuration of the exit sub-pupils in terms of achievable nulling ratio, that is characterized by perfect compensation of chromatic flux unbalance (the $A_R A_T$ and $A_T A_R$ exit beams). This intrinsic amplitude symmetry offers to the CCN a serious advantage with respect to the classical Mach-Zehnder configuration where the destructive port is used for nulling [8]. We also presented the first results of a laboratory experiment confirming the gain procured by the new CCN arrangement. So far the best obtained nulling ratio is equal to $1.8 \, 10^{-3}$, below digitization and readout noises levels. Additional work is required for evaluating the influence of the spectral bandwidth of the employed laser source and polarization effects at the cubes. It is also planned to improve the dynamic range of the camera and to eliminate the observed parasitic interferograms.

FH acknowledges funding help from the French "Action spécifique haute résolution angulaire" (ASHRA) managed by CNRS-INSU. He also thank his colleague B. Lazareff for careful reading of the manuscript.

---

[1] See e.g. the PERSEE experiment in Ref. [8] or Refs. [11-12] for mathematical treatment.

# REFERENCES

[1] A. Ferrari, R. Soummer, C. Aime, "An introduction to stellar coronagraphy," C. R. Phys. vol. 8, p. 277-287 (2007).

[2] R.N Bracewell, R.H. MacPhie "Searching for non solar planets," Icarus vol. 38, p. 136-147 (1979).

[3] A. Léger, J. M. Mariotti, B. Mennesson, M. Ollivier, J. L. Puget, D. Rouan, J. Schneider, "Could we search for primitive life on extrasolar planets in the near future ? The Darwin project," Icarus 123, p. 249-255 (1996).

[4] S. Seager, "Exoplanet transit spectroscopy and photometry," Space Science Reviews vol. 135, p. 345-354 (2008).

[5] F. Hénault, A. Spang, "Cheapest nuller in the world: Crossed beamsplitter cubes," Proceedings of the SPIE vol. 9146, n° 914604 (2014).

[6] F. Hénault, "Quantum physics and the beam splitter mystery," Proceedings of the SPIE vol. 9570, n° 95700Q (2015).

[7] A. Carlotti, G. Ricort, C. Aime, Y. El Azhari, R. Soummer, "Interferometric apodization of telescope apertures I. First laboratory results obtained using a Mach-Zehnder interferometer," Astronomy and Astrophysics vol. 477, p. 329-335 (2008).

[8] J.M. Le Duigou, J. Lozi, F. Cassaing, K. Houairi, B. Sorrente, J. Montri, S. Jacquinod, J.M. Reess, L. Pham, E. Lhome, T. Buey, F. Hénault, A. Marcotto, P. Girard, N. Mauclert, M. Barillot, V. Coudé du Foresto, M. Ollivier, "Final results of the PERSEE experiment," Proceedings of the SPIE vol. 8445, n° 844525 (2012).

[9] E. Ribak, S. G. Lipson, "Complex spatial coherence function: its measurement by means of a phase modulated shearing interferometer". Appl. Opt. vol. 20, p. 1102-1106 (1981).

[10] J. A. Ferrari, E. M. Frins, "Single-element interferometer," Optics Communications vol. 279, p. 235-239 (2007).

[11] F. Hénault, "Computing extinction maps of star nulling interferometers," Opt. Exp. 16, 4537-4546 (2008).

[12] F. Hénault, "Fine art of computing nulling interferometer maps," Proceedings of the SPIE 7013, n° 70131X (2008).